\documentclass[galaxies,article,submit,moreauthors]{Definitions/mdpi} 

\firstpage{1} 
\pubvolume{1}
\issuenum{1}
\articlenumber{0}
\pubyear{2024}
\copyrightyear{2024}
\datereceived{ } 
\daterevised{ } 
\dateaccepted{ } 
\datepublished{ } 
\hreflink{https://doi.org/} 

\Title{Spectroscopy of a sample of RV Tauri stars without IR excess}

\TitleCitation{Spectroscopy of a sample of RV Tauri stars without IR excess}

\Author{Kārlis Puķītis *\orcidA{} and Karina Korenika} 

\AuthorNames{Kārlis Puķītis and Karina Korenika}

\AuthorCitation{Puķītis, K.; Korenika, K.}

\address [1] {%
\quad Laser Centre, Faculty of Sciences and Technology, University of Latvia, Raiņa bulvāris 19, LV-1586 Rīga, Latvia; karlis.pukitis@lu.lv}

\corres{Correspondence: karlis.pukitis@lu.lv}

\abstract{We observed high-resolution optical spectra of 11 RV Tauri stars without IR excess, with the primary goal of searching for chemical depletion patterns. Using equivalent widths of absorption lines, we calculated photospheric parameters and chemical element abundances for five stars in the sample: HD 172810, V399 Cyg, AA Ari, V457 Cyg, and V894 Per. Only the abundance pattern of V457 Cyg suggests depletion. In the spectrum of this star, TiO lines are also observed in emission in addition to metal emissions. V457 Cyg is likely a binary system that was once surrounded by a circumbinary disc. In the spectrum of V894 Per, we find a set of spectral lines that appear to belong to another star, corroborating that it is an eclipsing variable rather than an RV Tauri star. The high overabundance of sodium may result from mass transfer within the binary system.}


\begin{document}

\section{Introduction}
The main characteristic of RV Tauri type stars is the presence of pulsation caused alternating deep and shallow minima in their light curves. Many of these variables possess a peculiar photospheric abundance pattern called depletion, where chemical elements with high dust condensation temperatures are systematically underabundant. This peculiarity is generally thought to arise from the accretion of gas from a dusty disc \cite{Oomen2019}. This surrounding structure causes IR excess in the spectral energy distribution of RV Tauri-type objects. Indeed, depletion has been observed primarily in stars that exhibit IR excess \cite{Gezer2015}. The reason why some RV Tauri stars surrounded by a dusty disc exhibit depletion, while others do not, may be related to the presence of a planet within the disc \cite{Kluska2022}. However, \cite{Gezer2015} found a few depleted RV Tauri stars without IR excess. 

From the perspective of stellar evolution, RV Tauri objects are thought to be in the post-asymptotic giant branch (post-AGB) or post-red giant branch (post-RGB) phase. Variables that possess IR excess are binaries, and the disc is a result of a binary interaction that terminated the AGB or RGB stage. The evolutionary status of RV Tauri stars without IR excess is less clear; these may be either single, low-luminosity post-AGB stars or binary post-RGB stars. In the latter case, it is thought that the disc has already dispersed \cite{Manick2018}. Little is known about the evolution of discs surrounding post-AGB and post-RGB stars. The longevity of these structures is supported by the fact that they contain large dust grains with high levels of crystallinity (see, e.g.,\cite{Gielen2008}). The evolution of such discs around evolved stars was modelled by \cite{Izzard2023}, who found that the main factor determining the disc's lifetime is the evolution of the luminous star in the central binary. When the surface of the central star reaches a temperature of around 3 $\times$ 10\textsuperscript{4} K, its X-ray flux causes rapid photoevaporation of the disc. However, significant mass loss from the disc in the form of wind has been observed for discs with central stars that have a surface temperature below 1 $\times$ 10\textsuperscript{4} K (see, e.g., \cite{Bujarrabal2018}).

RV Tauri stars whose discs have already dissipated could provide important clues about the evolution of these circumbinary structures. Finding depletion in a star without IR excess indicates that a disc was present at an earlier time. In this study, we present our spectroscopic observations of such RV Tauri variables and the results of their initial analysis.

\section{Observations}
We observed high-resolution spectra of 11 RV Tauri-type stars without IR excess, with the primary goal of searching for depletion patterns. These stars were selected from the compilation by \cite{Gezer2015}, focusing on the brightest objects that had not yet been studied via high-resolution spectroscopy. The observations were conducted using the high-resolution FIbre-fed Echelle Spectrograph (FIES; \cite{Telting2014}) at the Nordic Optical Telescope and the Vilnius University Echelle Spectrograph (VUES; \cite{Jurjenson2016}) at the 1.65-metre telescope at the Molėtai Astronomical Observatory. The observations are listed in Table~\ref{tab:observations}. Reduction of the observed spectra was conducted by using automatic pipelines of the respective observatories.

\begin{table}[H] 
\caption{High-resolution spectroscopic observations of RV Tauri stars without IR excess.\label{tab:observations}}
\begin{tabularx}{\textwidth}{CCCCCC}
\toprule
Object    & V     & Date        & Spectrograph & R     & Exp. time (min) \\
\midrule
\multirow[t]{3}{*}{V428 Aur} & \multirow[t]{3}{*}{6.87}  & 2021 Feb 19 & \multirow[t]{3}{*}{VUES} & 30000 & 135 \\
          &       & 2021 Mar 3  &                          & 30000 & 60              \\
     \vspace{0.1cm}     &     \vspace{0.1cm}  & 2022 Oct 12\vspace{0.1cm} &           \vspace{0.1cm}               & 30000\vspace{0.1cm} & 120\vspace{0.1cm}             \\
HD 172810\vspace{0.1cm} & 8.42\vspace{0.1cm}  & 2022 Sep 11\vspace{0.1cm} & FIES\vspace{0.1cm}               & 46000\vspace{0.1cm} & 11\vspace{0.1cm}              \\
V399 Cyg\vspace{0.1cm}  & 11.28\vspace{0.1cm} & 2022 Sep 11\vspace{0.1cm} & FIES\vspace{0.1cm}               & 46000\vspace{0.1cm} & 134\vspace{0.1cm}             \\
AA Ari\vspace{0.1cm}    & 8.69\vspace{0.1cm}  & 2022 Sep 12\vspace{0.1cm} & FIES \vspace{0.1cm}              & 46000\vspace{0.1cm} & 12\vspace{0.1cm}              \\
V360 Peg\vspace{0.1cm}  & 8.23\vspace{0.1cm}  & 2022 Sep 12\vspace{0.1cm} & FIES\vspace{0.1cm}               & 46000\vspace{0.1cm} & 8\vspace{0.1cm}               \\
HD 143352\vspace{0.1cm} & 9.33\vspace{0.1cm}  & 2023 Feb 27\vspace{0.1cm} & VUES\vspace{0.1cm}         & 30000\vspace{0.1cm} & 120\vspace{0.1cm}             \\
\multirow[t]{5}{*}{DZ UMa}   & \multirow[t]{5}{*}{11.14} & 2023 Feb 27 & \multirow[t]{5}{*}{VUES} & 30000 & 60  \\
          &       & 2023 Apr 22 &                          & 30000 & 180             \\
          &       & 2023 Apr 23 &                          & 30000 & 180             \\
          &       & 2023 Apr 24 &                          & 30000 & 60              \\
    \vspace{0.1cm}      &   \vspace{0.1cm}    & 2023 Apr 27\vspace{0.1cm} &            \vspace{0.1cm}              & 30000\vspace{0.1cm} & 60\vspace{0.1cm}              \\
V1673 Cyg\vspace{0.1cm} & 11.73\vspace{0.1cm} & 2023 Aug 17\vspace{0.1cm} & FIES\vspace{0.1cm}               & 25000\vspace{0.1cm} & 97\vspace{0.1cm}              \\
V457 Cyg\vspace{0.1cm}  & 11.99\vspace{0.1cm} & 2023 Aug 17\vspace{0.1cm} & FIES\vspace{0.1cm}               & 25000\vspace{0.1cm} & 123\vspace{0.1cm}             \\
V362 Aql\vspace{0.1cm}  & 11.79\vspace{0.1cm} & 2023 Aug 17\vspace{0.1cm} & FIES \vspace{0.1cm}              & 25000\vspace{0.1cm} & 103 \vspace{0.1cm}            \\
V894 Per  & 11.94 & 2023 Aug 17 & FIES               & 25000 & 118             \\
\bottomrule
\end{tabularx}
\end{table}

\begin{figure}[H]
\includegraphics[width=\textwidth, trim={1cm 0.85cm 1.95cm 1.5cm},clip]{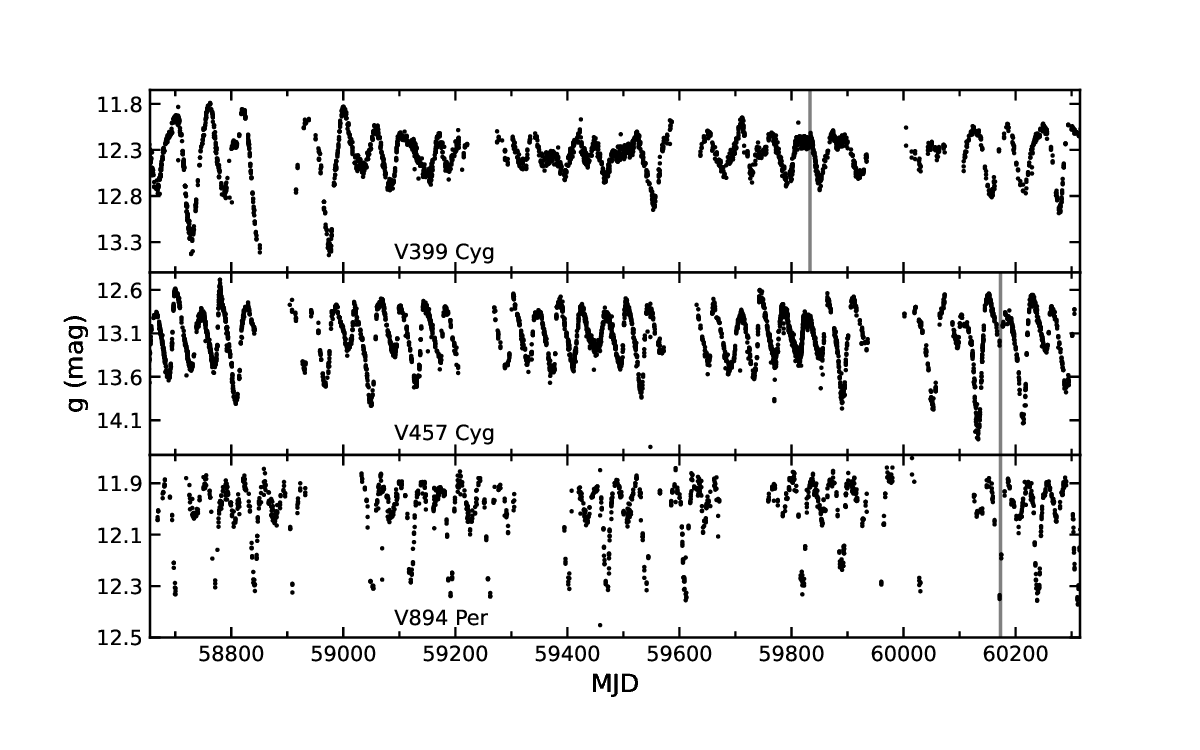}
\caption{The ASAS-SN light curves of V399 Cyg, V457 Cyg, and V894 Per. The grey vertical lines indicate the times of the spectroscopic observations.\label{fig:ASASSN}}
\label{depletion}
\end{figure}

Here, we present the initial results of the spectroscopic analysis for five of the observed stars: HD 172810, V399 Cyg, AA Ari, V457 Cyg, and V894 Per. We use the All-Sky Automated Survey for Supernovae (ASAS-SN; \cite{Shapee2014,Kochanek2017}) light curves (Figure ~\ref{fig:ASASSN}) to ascertain their RV Tauri nature and the pulsation phase that corresponds to the time the spectra were observed. The spectrum of V399 Cyg corresponds to the local light maximum just before dimming. However, close to the time of our spectroscopic observations, this star did not exhibit the typical alternating deep and shallow minima. Nevertheless, RV Tauri-like pulsation was observed for some time before MJD $\approx$ 59100. The spectrum of V457 Cyg corresponds to the shallow minimum in the light curve. This star also appears to often pulsate in a semiregular manner instead of in an RV Tauri fashion. The spectrum of V894 Per corresponds to the deep minimum in the light curve, where the alternating behaviour is clearly seen at all times. The ASAS-SN light curves of HD 172810 and AA Ari appear to be problematic; therefore, we cannot determine the pulsation phase that corresponds to our spectra and confirm the RV Tauri nature. None of the analysed stars show long-term mean magnitude variation in their light curves. We will present the spectroscopic analysis for the remainder of the observed stars in a future paper.

\unskip

\section{Analysis}
The reduced spectra are normalised by dividing them with a spline fitted to interactively placed continuum points. Line lists extracted from the Vienna Atomic Line Database (VALD; \cite{Piskunov1995,Kupka1999}) are used to identify spectral lines. Radial velocities are estimated using a large number of weak and medium-strength absorption lines in each spectrum and by cross-correlating the line profiles with their mirror profiles. The least blended lines are utilized to measure their equivalent widths (EWs), which are used to derive the chemical element abundances. The calculations are performed using the SPECTRUM code \cite{Gray1994} and plane-parallel LTE model atmospheres from the ATLAS grid \cite{Kurucz2005}. For V399 Cyg, the iSpec code \cite{Blanco-Cuaresma2014} is used to interpolate the model atmospheres. The source of the atomic data is VALD. Continuum normalisation, along with the measurement of radial velocities and EWs, is performed using the DECH software\footnote{\url{https://gazinur.crao.ru/Download.html}}.

The photospheric parameters are determined by employing the standard method of excitation and ionization balance for iron lines. The effective temperature is found by minimising the dependence of iron abundances on the excitation potential of the respective lines. First, the abundances derived from individual Fe I lines (and additionally, Fe II lines in the case of V894 Per) are plotted as a function of the excitation potential. If a trend between the abundances and excitation potentials is observed, the abundances are recalculated by selecting a model atmosphere with a different effective temperature. This process continues until a model atmosphere that results in no trend is found. Subsequently,
the surface gravity is determined by selecting a model atmosphere that yields the same iron abundances derived from Fe I and Fe II lines. In the case of AA Ari, we are unable to find suitable Fe II lines; therefore, the surface gravity is highly uncertain. It is approximately estimated by synthesising several regions of the spectrum and attempting to find the best fit with the observations. The microturbulent velocity is derived in a manner similar to that used for the effective temperature, with the difference being that we minimise the trend with the EW instead of the excitation potential.

\section{Results and Discussion}
The results for the photospheric parameters and radial velocities are presented in Table~\ref{tab:param}, while the derived abundances are presented in Table~\ref{tab:abundances}. 

\begin{table}[H] 
\caption{Photospheric parameters and heliocentric radial velocities of the analysed stars.\label{tab:param}}
\begin{tabularx}{\textwidth}{CCCCC}
\toprule
Object & $T_{eff}$	& log\,$g$ & ${V_{mic}}$ & ${V_{r}}$\\
	& (K)	& & (km/s) & (km/s)\\
\midrule
HD 172810 & 4250   & 2.5   & 2              & -4.1       \\
V399 Cyg  & 4150   & 0     & 2.7            & -42.6      \\
AA Ari    & 4750   & 2\textsuperscript{a}     & 2.8            & -43.2      \\
V457 Cyg  & 5750   & 0.5   & 5              & -53.4      \\
V894 Per  & 8000\textsuperscript{b}   & 1\textsuperscript{b}     & 6.8\textsuperscript{b}            & -64.3\textsuperscript{c}      \\
\bottomrule
\end{tabularx}
\noindent{\footnotesize{\textsuperscript{a} Uncertain due to lack of suitable Fe II lines.}}
\noindent{\footnotesize{\textsuperscript{b} Uncertain due to another star affecting the spectrum.}}
\noindent{\footnotesize{\textsuperscript{c} Another set of spectral lines at around -137 km/s visible in the spectrum.}}
\end{table}

\begin{table}[H]
\caption{Photospheric abundances of the analysed stars.\label{tab:abundances}}
	\begin{adjustwidth}{-\extralength}{0cm}
		\begin{tabularx}{\fulllength}{C|CCC|CCC|CCC|CCC|CCC}
			\toprule
             & \multicolumn{3}{c|}{HD 172810} & \multicolumn{3}{c|}{V399 Cyg} & \multicolumn{3}{c|}{AA Ari} & \multicolumn{3}{c|}{V457 Cyg} & \multicolumn{3}{c}{V894 Per}\\
			Ion & [X/H]	& N & $\sigma$ & [X/H]	& N & $\sigma$ &[X/H]	& N & $\sigma$ &[X/H]	& N & $\sigma$ &[X/H]	& N & $\sigma$\\
			\midrule
C I   &       &     &      &       &    &      &       &    &      & -0.51 & 5   & 0.13 &      &    &      \\
O I   & 0.49  & 1   &      & -0.63 & 2  & 0.04 &       &    &      &       &     &      &      &    &      \\
Na I  & 0.25  & 2   & 0.02 &       &    &      & 0.89  & 1  &      & -0.91 & 1   &      & 1.63 & 3  & 0.40 \\
Mg I  & -0.03 & 4   & 0.15 & -0.82 & 1  &      & -0.19 & 1  &      & -0.48 & 2   & 0.03 & 0.32 & 4  & 0.06 \\
Mg II &       &     &      &       &    &      &       &    &      &       &     &      & 0.27 & 1  &      \\
Al I  & 0.20  & 4   & 0.24 &       &    &      &       &    &      &       &     &      &      &    &      \\
Si I  & 0.15  & 7   & 0.14 &       &    &      & 0.13  & 1  &      & -0.27 & 14  & 0.14 &      &    &      \\
Si II &       &     &      &       &    &      &       &    &      &       &     &      & 0.42 & 1  &      \\
S I   &       &     &      &       &    &      &       &    &      & -0.23 & 4   & 0.10 & 0.86 & 1  &      \\
K I   &       &     &      &       &    &      &       &    &      & -1.03 & 2   & 0.09 &      &    &      \\
Ca I  & -0.35 & 10  & 0.20 & -1.42 & 7  & 0.07 &       &    &      & -1.55 & 4   & 0.12 &      &    &      \\
Ca II &       &     &      &       &    &      &       &    &      &       &     &      & 0.77 & 1  &      \\
Sc I  & 0.23  & 6   & 0.12 &       &    &      & 1.46  & 2  & 0.47 &       &     &      &      &    &      \\
Sc II & 0.08  & 2   & 0.16 & -1.81 & 1  &      & -0.43 & 1  &      &       &     &      & 0.30 & 4  & 0.10 \\
Ti I  & 0.24  & 70  & 0.25 & -1.36 & 7  & 0.10 & 0.79  & 10 & 0.22 &       &     &      &      &    &      \\
Ti II & -0.07 & 1   &      & -1.18 & 6  & 0.12 &       &    &      & -1.41 & 12  & 0.13 & 0.39 & 14 & 0.15 \\
V I   & 0.56  & 42  & 0.28 & -1.63 & 7  & 0.13 & 0.92  & 8  & 0.21 &       &     &      &      &    &      \\
V II  &       &     &      &       &    &      &       &    &      &       &     &      & 0.32 & 2  & 0.09 \\
Cr I  & 0.10  & 19  & 0.13 & -1.74 & 6  & 0.11 & 0.59  & 1  &      &       &     &      &      &    &      \\
Cr II & -0.10 & 1   &      &       &    &      &       &    &      & -1.62 & 2   & 0.05 & 0.32 & 15 & 0.13 \\
Mn I  & 0.12  & 2   & 0.16 & -1.84 & 4  & 0.35 & 1.09  & 2  & 0.71 & -1.30 & 2   & 0.17 &      &    &      \\
Fe I  & 0.01  & 160 & 0.24 & -1.47 & 54 & 0.14 & 0.07  & 35 & 0.24 & -0.91 & 122 & 0.19 & 0.40 & 21 & 0.19 \\
Fe II & -0.04 & 2   & 0.16 & -1.52 & 6  & 0.13 &       &    &      & -1.02 & 24  & 0.17 & 0.26 & 26 & 0.16 \\
Co I  & 0.28  & 11  & 0.14 & -1.42 & 5  & 0.20 &       &    &      &       &     &      &      &    &      \\
Ni I  & 0.08  & 6   & 0.18 & -1.16 & 18 & 0.24 & -0.49 & 1  &      & -0.81 & 14  & 0.07 & 0.37 & 1  &      \\
Ni II &       &     &      &       &    &      &       &    &      & -0.89 & 1   &      & 0.32 & 1  &      \\
Cu I  &       &     &      & -1.23 & 2  & 0.31 &       &    &      &       &     &      &      &    &      \\
Zn I  &       &     &      & -1.42 & 1  &      &       &    &      & -0.74 & 2   & 0.00 &      &    &      \\
Sr I  & 0.39  & 1   &      &       &    &      &       &    &      &       &     &      &      &    &      \\
Y I   & 0.25  & 2   & 0.01 &       &    &      & 0.77  & 1  &      &       &     &      &      &    &      \\
Y II  & 0.28  & 2   & 0.13 & -1.70 & 2  & 0.06 & 0.49  & 1  &      &       &     &      & 0.37 & 5  & 0.18 \\
Zr I  & 0.52  & 9   & 0.28 & -1.34 & 1  &      & 0.71  & 1  &      &       &     &      & 0.51 & 5  & 0.10 \\
Nb I  & 0.78  & 2   & 0.39 &       &    &      &       &    &      &       &     &      &      &    &      \\
Mo I  & 0.35  & 5   & 0.10 &       &    &      &       &    &      &       &     &      &      &    &      \\
Ba II &       &     &      &       &    &      &       &    &      &       &     &      & 0.41 & 1  &      \\
La II & 0.70  & 6   & 0.33 &       &    &      & 0.45  & 1  &      &       &     &      &      &    &      \\
Ce II & 0.52  & 1   &      & -1.72 & 2  & 0.15 &       &    &      &       &     &      &      &    &      \\
Pr II & 0.53  & 3   & 0.34 &       &    &      &       &    &      &       &     &      &      &    &      \\
Nd II & 0.39  & 5   & 0.14 & -1.36 & 4  & 0.24 & 0.70  & 2  & 0.28 &       &     &      &      &    &      \\
Sm II &       &     &      & -1.28 & 5  & 0.13 &       &    &      &       &     &      &      &    &      \\

			\bottomrule
		\end{tabularx}
	\end{adjustwidth}
	\noindent{\footnotesize{*Abundances from \cite{Asplund2009} are used as a reference for solar values.}}
\end{table}

\subsection{HD 172810, V399 Cyg, and AA Ari}
The derived abundances of HD 172810, V399 Cyg, and AA Ari show no evidence of depletion. Neutron-capture elements in HD 172810 appear to have a slight overabundance ([n/H] = 0.4-0.5), while lighter elements have approximately solar abundances. The enhancement of heavy elements is unlikely to result from intrinsic nucleosynthesis due to the star's low luminosity \cite{BodiKiss2019}. All elements in V399 Cyg have approximately the same abundance ([X/H] $\approx$ -1.5), except for oxygen and magnesium. The lines in the spectrum of AA Ari are broad; the average full width at half maximum (FWHM) corresponds to 28 km/s on the velocity scale (FWHM $\approx$ 10 km/s for HD 172810 and V399 Cyg). This contributes to the large uncertainty in the abundances, which are calculated using more than two lines only for titanium, vanadium, and iron. While iron has approximately solar abundance, titanium and vanadium have [X/H] = 0.8-0.9. 

\subsection{V457 Cyg}
\begin{figure}[H]
\includegraphics[width=\linewidth, trim={-1.25cm 0.8cm -1.25cm 0.8cm},clip]{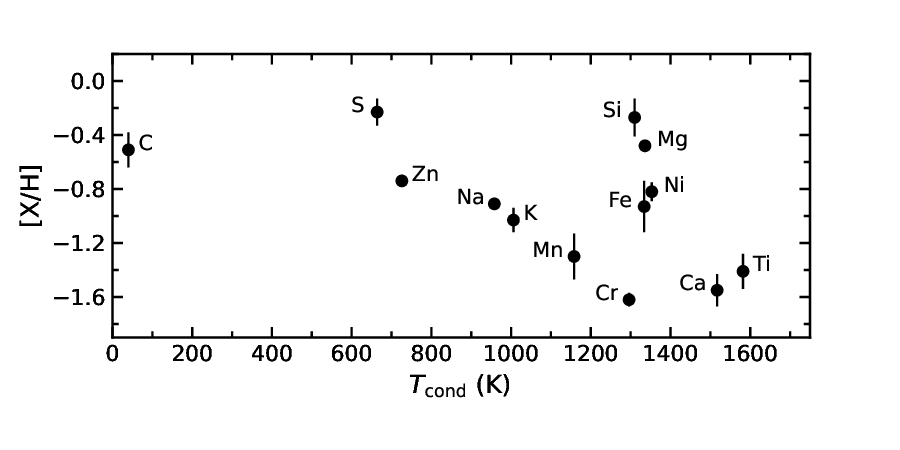}
\caption{Photospheric abundances of V457 Cyg as a function of condensation temperature, using values from \cite{Lodders2003}. Vertical lines indicate the standard deviation of the calculated abundances.\label{fig:depletion}}
\end{figure}

\begin{figure}[H]
\includegraphics[width=\linewidth, trim={-1.25cm 0.8cm -1.25cm 0.8cm},clip]{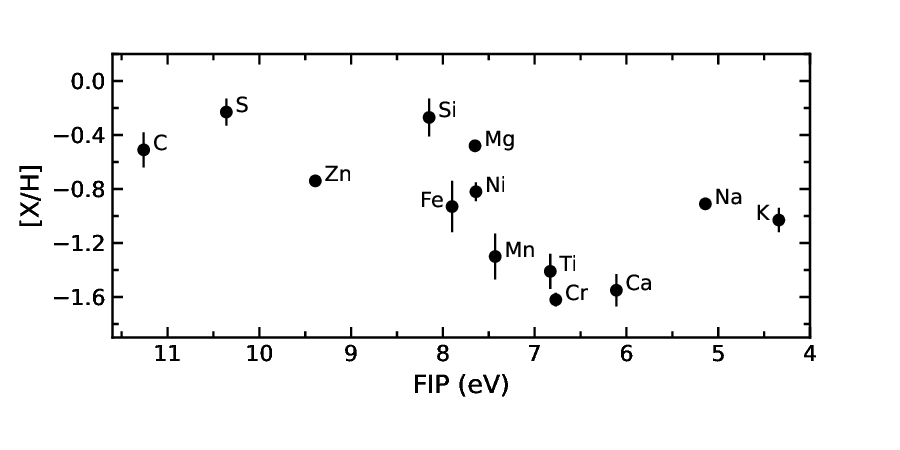}
\caption{Photospheric abundances of V457 Cyg as a function of the FIP.
\label{fig:FIP}}
\end{figure}

\begin{figure}[H]
\begin{adjustwidth}{-\extralength}{0cm}
\includegraphics[width=\linewidth, trim={2.15cm 0.3cm 2.5cm 0.45cm},clip]{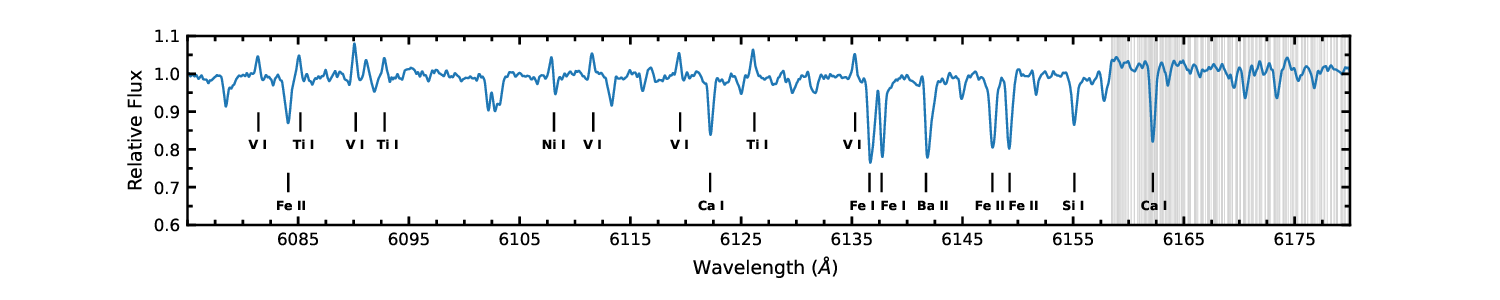}
\end{adjustwidth}
\caption{Region of the spectrum of V457 Cyg where emission lines are clearly visible. The strongest absorption and emission lines are identified with black vertical lines, while grey vertical lines correspond to the wavelengths of the TiO spectral lines.\label{fig:emisija}}
\end{figure}

In the spectrum of V457 Cyg, the absorption lines are also broad---FWHM = 23 km/s on average. Numerous weak metal emission lines are visible in the spectrum, identified as low-excitation lines of mostly vanadium and titanium. We also identify TiO molecule lines in emission (Figure~\ref{fig:emisija}). The abundance pattern of V457 Cyg indicates depletion, as the abundance ratios [Zn/Ti] and [S/Ti], which are used to quantify the depletion \cite{Gezer2015}, are 0.67 and 1.18, respectively. Therefore, this star is very likely a binary,  and its surface chemical composition results from the accretion from a circumbinary disc that has already dissipated. V457 Cyg is the fourth RV Tauri star without IR excess known to be depleted, with the other three being SS Gem, AZ Sgr, and EQ Cas \cite{Gezer2015}. However, when examining the abundances of all the elements in the photosphere of V457 Cyg (Figure~\ref{fig:depletion}), a more complicated picture emerges. The abundances of silicon and magnesium appear too high to be consistent with the depletion scenario. These can be explained by the initial composition of the star being alpha-enhanced by around 0.5 dex. A similar abundance pattern was found in EQ Cas, which was explained as a consequence of the so-called first ionization potential (FIP) effect, in which stellar winds preferentially pick up ions over atoms\cite{Giridhar2005,Rao2005}. This results in elements with low FIP having lower abundances. However, in the case of V457 Cyg, sodium and potassium abundances appear to be inconsistent with the FIP effect (Figure~\ref{fig:FIP}).

\unskip

\unskip

\unskip

\begin{figure}[H]
\begin{adjustwidth}{-\extralength}{0cm}
\includegraphics[width=\linewidth, trim={2.15cm 0.3cm 2.5cm 0.45cm},clip]{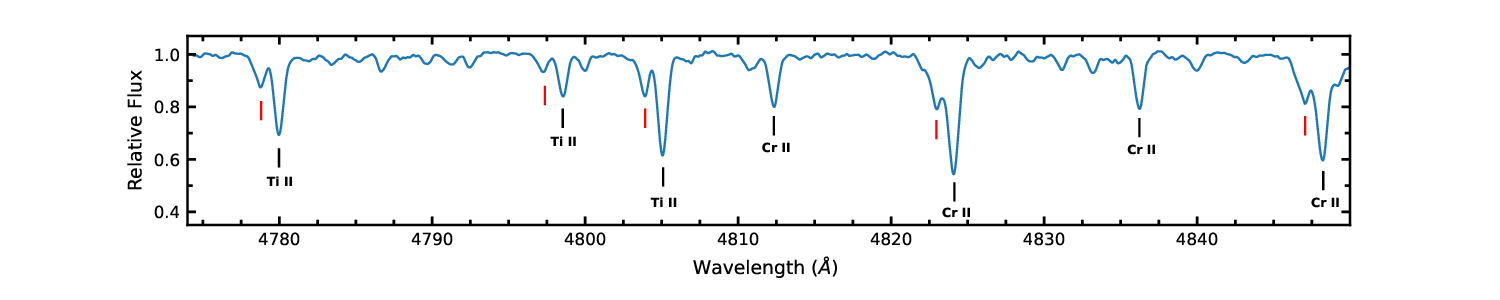}
\end{adjustwidth}
\caption{Region of the spectrum of V894 Per where two sets of absorption lines are clearly visible. The most intense spectral lines, corresponding to a radial velocity of -64.3 km/s, are identified with black vertical lines. To the left of these lines, the same absorptions, corresponding to a radial velocity of approximately -137 km/s, are marked.\label{fig:binary}}
\end{figure}   
\unskip

\subsection{V894 Per}
The spectral lines of V894 Per are very broad---FWHM $\approx$ 38 km/s. The derived abundances show [X/H] $\approx$ 0.4 on average. The abundance of sodium is considerably larger---[Na/H] $\approx$ 1.6. Some of the strongest identified ion lines are accompanied by a weaker component that is shifted by around 73 km/s to the short-wavelength side, suggesting that the star is a binary. This supports the idea that V894 Per could be an eclipsing variable instead of an RV Tauri star \cite{Khruslov2006, Kazarovets2011, Nere2024}. Since the spectrum is contaminated by the light from the other component of the binary, our results for the photospheric parameters and abundances should be regarded as uncertain. Nevertheless, the overabundance of sodium is most likely real and may be a consequence of mass transfer within the binary system.
\cite{Ripepi2024}.

\vspace{6pt} 

\authorcontributions{Conceptualisation, visualisation, writing---original draft preparation, writing---review and editing, K.P.; formal analysis, K.P and K.K. All authors have read and agreed to the published version of the manuscript.}

\acknowledgments{We acknowledge funding from the Latvian Council of Science under the project “Advanced Spectroscopic Methods and Tools for the Study of Evolved Stars,” project No. lzp-flpp-2020/1-0088.
The research leading to these results received funding from the European Community's Horizon 2020 Programme (H2020/2021-2024) under grant agreement number 101008324 (ChETEC-INFRA). Based on observations made with the Nordic Optical Telescope, owned in collaboration by the University of Turku and Aarhus University, and operated jointly by Aarhus University, the University of Turku, and the University of Oslo, representing Denmark, Finland, and Norway, as well as the University of Iceland and Stockholm University at the Observatorio del Roque de los Muchachos, La Palma, Spain, of the Instituto de Astrofisica de Canarias. This work used the VALD database, maintained by Uppsala University, the Institute of Astronomy RAS in Moscow, and the University of Vienna.}

\conflictsofinterest{The authors declare no conflicts of interest.}

\begin{adjustwidth}{-\extralength}{0cm}

\reftitle{References}
\bibliography{atsauces.bib}

\begin{thebibliography}{999}

\bibitem[{Oomen} et~al.(2019){Oomen}, {Van Winckel}, {Pols}, and
  {Nelemans}]{Oomen2019}
{Oomen}, G.M.; {Van Winckel}, H.; {Pols}, O.; {Nelemans}, G.
\newblock {Modelling depletion by re-accretion of gas from a dusty disc in
  post-AGB stars}.
\newblock {\em Astronomy \& Astrophysics} {\bf 2019}, {\em 629},~A49,
  \href{http://arxiv.org/abs/1908.01788}{{\normalfont
  [arXiv:astro-ph.SR/1908.01788]}}.
\newblock {\url{https://doi.org/10.1051/0004-6361/201935853}}.

\bibitem[{Gezer} et~al.(2015){Gezer}, {Van Winckel}, {Bozkurt}, {De Smedt},
  {Kamath}, {Hillen}, and {Manick}]{Gezer2015}
{Gezer}, I.; {Van Winckel}, H.; {Bozkurt}, Z.; {De Smedt}, K.; {Kamath}, D.;
  {Hillen}, M.; {Manick}, R.
\newblock {The WISE view of RV Tauri stars}.
\newblock {\em Monthly Notices of the Royal Astronomical Society} {\bf 2015},
  {\em 453},~133--146,  \href{http://arxiv.org/abs/1507.04175}{{\normalfont
  [arXiv:astro-ph.SR/1507.04175]}}.
\newblock {\url{https://doi.org/10.1093/mnras/stv1627}}.

\bibitem[{Kluska} et~al.(2022){Kluska}, {Van Winckel}, {Copp{\'e}e}, {Oomen},
  {Dsilva}, {Kamath}, {Bujarrabal}, and {Min}]{Kluska2022}
{Kluska}, J.; {Van Winckel}, H.; {Copp{\'e}e}, Q.; {Oomen}, G.M.; {Dsilva}, K.;
  {Kamath}, D.; {Bujarrabal}, V.; {Min}, M.
\newblock {A population of transition disks around evolved stars: Fingerprints
  of planets. Catalog of disks surrounding Galactic post-AGB binaries}.
\newblock {\em Astronomy \& Astrophysics} {\bf 2022}, {\em 658},~A36,
  \href{http://arxiv.org/abs/2201.13155}{{\normalfont
  [arXiv:astro-ph.EP/2201.13155]}}.
\newblock {\url{https://doi.org/10.1051/0004-6361/202141690}}.

\bibitem[{Manick} et~al.(2018){Manick}, {Van Winckel}, {Kamath}, {Sekaran}, and
  {Kolenberg}]{Manick2018}
{Manick}, R.; {Van Winckel}, H.; {Kamath}, D.; {Sekaran}, S.; {Kolenberg}, K.
\newblock {The evolutionary nature of RV Tauri stars in the SMC and LMC}.
\newblock {\em Astronomy \& Astrophysics} {\bf 2018}, {\em 618},~A21,
  \href{http://arxiv.org/abs/1806.08210}{{\normalfont
  [arXiv:astro-ph.SR/1806.08210]}}.
\newblock {\url{https://doi.org/10.1051/0004-6361/201833130}}.

\bibitem[{Gielen} et~al.(2008){Gielen}, {van Winckel}, {Min}, {Waters}, and
  {Lloyd Evans}]{Gielen2008}
{Gielen}, C.; {van Winckel}, H.; {Min}, M.; {Waters}, L.B.F.M.; {Lloyd Evans},
  T.
\newblock {SPITZER survey of dust grain processing in stable discs around
  binary post-AGB stars}.
\newblock {\em Astronomy \& Astrophysics} {\bf 2008}, {\em 490},~725--735,
  \href{http://arxiv.org/abs/0809.2505}{{\normalfont
  [arXiv:astro-ph/0809.2505]}}.
\newblock {\url{https://doi.org/10.1051/0004-6361:200810053}}.

\bibitem[{Izzard} and {Jermyn}(2023)]{Izzard2023}
{Izzard}, R.G.; {Jermyn}, A.S.
\newblock {Circumbinary discs for stellar population models}.
\newblock {\em Monthly Notices of the Royal Astronomical Society} {\bf 2023},
  {\em 521},~35--50,  \href{http://arxiv.org/abs/2401.14315}{{\normalfont
  [arXiv:astro-ph.SR/2401.14315]}}.
\newblock {\url{https://doi.org/10.1093/mnras/stac2899}}.

\bibitem[{Bujarrabal} et~al.(2018){Bujarrabal}, {Castro-Carrizo}, {Van
  Winckel}, {Alcolea}, {S{\'a}nchez Contreras}, {Santander-Garc{\'\i}a}, and
  {Hillen}]{Bujarrabal2018}
{Bujarrabal}, V.; {Castro-Carrizo}, A.; {Van Winckel}, H.; {Alcolea}, J.;
  {S{\'a}nchez Contreras}, C.; {Santander-Garc{\'\i}a}, M.; {Hillen}, M.
\newblock {High-resolution observations of IRAS 08544-4431. Detection of a disk
  orbiting a post-AGB star and of a slow disk wind}.
\newblock {\em Astronomy \& Astrophysics} {\bf 2018}, {\em 614},~A58,
  \href{http://arxiv.org/abs/1802.04019}{{\normalfont
  [arXiv:astro-ph.SR/1802.04019]}}.
\newblock {\url{https://doi.org/10.1051/0004-6361/201732422}}.

\bibitem[{Telting} et~al.(2014){Telting}, {Avila}, {Buchhave}, {Frandsen},
  {Gandolfi}, {Lindberg}, {Stempels}, {Prins}, and {NOT staff}]{Telting2014}
{Telting}, J.H.; {Avila}, G.; {Buchhave}, L.; {Frandsen}, S.; {Gandolfi}, D.;
  {Lindberg}, B.; {Stempels}, H.C.; {Prins}, S.; {NOT staff}.
\newblock {FIES: The high-resolution Fiber-fed Echelle Spectrograph at the
  Nordic Optical Telescope}.
\newblock {\em Astronomische Nachrichten} {\bf 2014}, {\em 335},~41.
\newblock {\url{https://doi.org/10.1002/asna.201312007}}.

\bibitem[{Jurgenson} et~al.(2016){Jurgenson}, {Fischer}, {McCracken}, {Sawyer},
  {Giguere}, {Szymkowiak}, {Santoro}, and {Muller}]{Jurjenson2016}
{Jurgenson}, C.; {Fischer}, D.; {McCracken}, T.; {Sawyer}, D.; {Giguere}, M.;
  {Szymkowiak}, A.; {Santoro}, F.; {Muller}, G.
\newblock {Design and Construction of VUES: The Vilnius University Echelle
  Spectrograph}.
\newblock {\em Journal of Astronomical Instrumentation} {\bf 2016}, {\em
  5},~1650003--239,  \href{http://arxiv.org/abs/1601.06024}{{\normalfont
  [arXiv:astro-ph.IM/1601.06024]}}.
\newblock {\url{https://doi.org/10.1142/S2251171716500033}}.

\bibitem[{Shappee} et~al.(2014){Shappee}, {Prieto}, {Grupe}, {Kochanek},
  {Stanek}, {De Rosa}, {Mathur}, {Zu}, {Peterson}, {Pogge}, {Komossa}, {Im},
  {Jencson}, {Holoien}, {Basu}, {Beacom}, {Szczygie{\l}}, {Brimacombe},
  {Adams}, {Campillay}, {Choi}, {Contreras}, {Dietrich}, {Dubberley},
  {Elphick}, {Foale}, {Giustini}, {Gonzalez}, {Hawkins}, {Howell}, {Hsiao},
  {Koss}, {Leighly}, {Morrell}, {Mudd}, {Mullins}, {Nugent}, {Parrent},
  {Phillips}, {Pojmanski}, {Rosing}, {Ross}, {Sand}, {Terndrup}, {Valenti},
  {Walker}, and {Yoon}]{Shapee2014}
{Shappee}, B.J.; {Prieto}, J.L.; {Grupe}, D.; {Kochanek}, C.S.; {Stanek}, K.Z.;
  {De Rosa}, G.; {Mathur}, S.; {Zu}, Y.; {Peterson}, B.M.; {Pogge}, R.W.;
  et~al.
\newblock {The Man behind the Curtain: X-Rays Drive the UV through NIR
  Variability in the 2013 Active Galactic Nucleus Outburst in NGC 2617}.
\newblock {\em The Astrophysical Journal} {\bf 2014}, {\em 788},~48,
  \href{http://arxiv.org/abs/1310.2241}{{\normalfont
  [arXiv:astro-ph.HE/1310.2241]}}.
\newblock {\url{https://doi.org/10.1088/0004-637X/788/1/48}}.

\bibitem[{Kochanek} et~al.(2017){Kochanek}, {Shappee}, {Stanek}, {Holoien},
  {Thompson}, {Prieto}, {Dong}, {Shields}, {Will}, {Britt}, {Perzanowski}, and
  {Pojma{\'n}ski}]{Kochanek2017}
{Kochanek}, C.S.; {Shappee}, B.J.; {Stanek}, K.Z.; {Holoien}, T.W.S.;
  {Thompson}, T.A.; {Prieto}, J.L.; {Dong}, S.; {Shields}, J.V.; {Will}, D.;
  {Britt}, C.;  et~al.
\newblock {The All-Sky Automated Survey for Supernovae (ASAS-SN) Light Curve
  Server v1.0}.
\newblock {\em Publications of the Astronomical Society of the Pacific} {\bf
  2017}, {\em 129},~104502,
  \href{http://arxiv.org/abs/1706.07060}{{\normalfont
  [arXiv:astro-ph.SR/1706.07060]}}.
\newblock {\url{https://doi.org/10.1088/1538-3873/aa80d9}}.

\bibitem[{Piskunov} et~al.(1995){Piskunov}, {Kupka}, {Ryabchikova}, {Weiss},
  and {Jeffery}]{Piskunov1995}
{Piskunov}, N.E.; {Kupka}, F.; {Ryabchikova}, T.A.; {Weiss}, W.W.; {Jeffery},
  C.S.
\newblock {VALD: The Vienna Atomic Line Data Base.}
\newblock {\em Astronomy and Astrophysics Supplement} {\bf 1995}, {\em
  112},~525.

\bibitem[{Kupka} et~al.(1999){Kupka}, {Piskunov}, {Ryabchikova}, {Stempels},
  and {Weiss}]{Kupka1999}
{Kupka}, F.; {Piskunov}, N.; {Ryabchikova}, T.A.; {Stempels}, H.C.; {Weiss},
  W.W.
\newblock {VALD-2: Progress of the Vienna Atomic Line Data Base}.
\newblock {\em Astronomy and Astrophysics Supplement} {\bf 1999}, {\em
  138},~119--133.
\newblock {\url{https://doi.org/10.1051/aas:1999267}}.

\bibitem[{Gray} and {Corbally}(1994)]{Gray1994}
{Gray}, R.O.; {Corbally}, C.J.
\newblock {The Calibration of MK Spectral Classes Using Spectral Synthesis. I.
  The Effective Temperature Calibration of Dwarf Stars}.
\newblock {\em Astronomical Journal} {\bf 1994}, {\em 107},~742.
\newblock {\url{https://doi.org/10.1086/116893}}.

\bibitem[{Kurucz}(2005)]{Kurucz2005}
{Kurucz}, R.L.
\newblock {ATLAS12, SYNTHE, ATLAS9, WIDTH9, et cetera}.
\newblock {\em Memorie della Societa Astronomica Italiana Supplementi} {\bf
  2005}, {\em 8},~14.

\bibitem[{Blanco-Cuaresma} et~al.(2014){Blanco-Cuaresma}, {Soubiran}, {Heiter},
  and {Jofr{\'e}}]{Blanco-Cuaresma2014}
{Blanco-Cuaresma}, S.; {Soubiran}, C.; {Heiter}, U.; {Jofr{\'e}}, P.
\newblock {Determining stellar atmospheric parameters and chemical abundances
  of FGK stars with iSpec}.
\newblock {\em Astronomy \& Astrophysics} {\bf 2014}, {\em 569},~A111,
  \href{http://arxiv.org/abs/1407.2608}{{\normalfont
  [arXiv:astro-ph.IM/1407.2608]}}.
\newblock {\url{https://doi.org/10.1051/0004-6361/201423945}}.

\bibitem[{Asplund} et~al.(2009){Asplund}, {Grevesse}, {Sauval}, and
  {Scott}]{Asplund2009}
{Asplund}, M.; {Grevesse}, N.; {Sauval}, A.J.; {Scott}, P.
\newblock {The Chemical Composition of the Sun}.
\newblock {\em Annual Review of Astronomy \& Astrophysics} {\bf 2009}, {\em
  47},~481--522,  \href{http://arxiv.org/abs/0909.0948}{{\normalfont
  [arXiv:astro-ph.SR/0909.0948]}}.
\newblock {\url{https://doi.org/10.1146/annurev.astro.46.060407.145222}}.

\bibitem[{B{\'o}di} and {Kiss}(2019)]{BodiKiss2019}
{B{\'o}di}, A.; {Kiss}, L.L.
\newblock {Physical Properties of Galactic RV Tauri Stars from Gaia DR2 Data}.
\newblock {\em The Astrophysical Journal} {\bf 2019}, {\em 872},~60,
  \href{http://arxiv.org/abs/1901.01409}{{\normalfont
  [arXiv:astro-ph.SR/1901.01409]}}.
\newblock {\url{https://doi.org/10.3847/1538-4357/aafc24}}.

\bibitem[{Giridhar} et~al.(2005){Giridhar}, {Lambert}, {Reddy}, {Gonzalez}, and
  {Yong}]{Giridhar2005}
{Giridhar}, S.; {Lambert}, D.L.; {Reddy}, B.E.; {Gonzalez}, G.; {Yong}, D.
\newblock {Abundance Analyses of Field RV Tauri Stars. VI. An Extended Sample}.
\newblock {\em The Astrophysical Journal} {\bf 2005}, {\em 627},~432--445,
  \href{http://arxiv.org/abs/astro-ph/0503344}{{\normalfont
  [arXiv:astro-ph/astro-ph/0503344]}}.
\newblock {\url{https://doi.org/10.1086/430265}}.

\bibitem[{Rao} and {Reddy}(2005)]{Rao2005}
{Rao}, N.K.; {Reddy}, B.E.
\newblock {High-resolution spectroscopy of the high galactic latitude RV Tauri
  star CE Virginis}.
\newblock {\em Monthly Notices of the Royal Astronomical Society} {\bf 2005},
  {\em 357},~235--241,
  \href{http://arxiv.org/abs/astro-ph/0411481}{{\normalfont
  [arXiv:astro-ph/astro-ph/0411481]}}.
\newblock {\url{https://doi.org/10.1111/j.1365-2966.2005.08636.x}}.

\bibitem[{Lodders}(2003)]{Lodders2003}
{Lodders}, K.
\newblock {Solar System Abundances and Condensation Temperatures of the
  Elements}.
\newblock {\em The Astrophysical Journal} {\bf 2003}, {\em 591},~1220--1247.
\newblock {\url{https://doi.org/10.1086/375492}}.

\bibitem[{Khruslov}(2006)]{Khruslov2006}
{Khruslov}, A.V.
\newblock {Tyc 3706 00485 1}.
\newblock {\em Peremennye Zvezdy Prilozhenie} {\bf 2006}, {\em 6},~23.

\bibitem[{Kazarovets} et~al.(2011){Kazarovets}, {Samus}, {Durlevich},
  {Kireeva}, and {Pastukhova}]{Kazarovets2011}
{Kazarovets}, E.V.; {Samus}, N.N.; {Durlevich}, O.V.; {Kireeva}, N.N.;
  {Pastukhova}, E.N.
\newblock {The 80th Name-List of Variable Stars. Part I - RA 0h to 6h}.
\newblock {\em Information Bulletin on Variable Stars} {\bf 2011}, {\em
  5969},~1.

\bibitem[{Nere} et~al.(2024){Nere}, {Montez}, and {S{\'a}nchez-Maes}]{Nere2024}
{Nere}, R.N.; {Montez}, R., J.; {S{\'a}nchez-Maes}, S.
\newblock {An Audit of the Light Curves of RV Tau Variable Stars in the ASAS-SN
  Database}.
\newblock {\em Journal of the American Association of Variable Star Observers}
  {\bf 2024}, {\em 52},~34,
  \href{http://arxiv.org/abs/2408.13309}{{\normalfont
  [arXiv:astro-ph.SR/2408.13309]}}.
\newblock {\url{https://doi.org/10.48550/arXiv.2408.13309}}.

\bibitem[{Ripepi} et~al.(2024){Ripepi}, {Catanzaro}, {Trentin}, {Straniero},
  {Mucciarelli}, {Marconi}, {Bhardwaj}, {Fiorentino}, {Monelli}, {Storm}, {De
  Somma}, {Leccia}, {Molinaro}, {Musella}, and {Sicignano}]{Ripepi2024}
{Ripepi}, V.; {Catanzaro}, G.; {Trentin}, E.; {Straniero}, O.; {Mucciarelli},
  A.; {Marconi}, M.; {Bhardwaj}, A.; {Fiorentino}, G.; {Monelli}, M.; {Storm},
  J.;  et~al.
\newblock {First spectroscopic investigation of anomalous Cepheid variables}.
\newblock {\em Astronomy \& Astrophysics} {\bf 2024}, {\em 682},~A1,
  \href{http://arxiv.org/abs/2310.20503}{{\normalfont
  [arXiv:astro-ph.GA/2310.20503]}}.
\newblock {\url{https://doi.org/10.1051/0004-6361/202347991}}.

\end{thebibliography}

\PublishersNote{}
\end{adjustwidth}
\end{document}